\documentclass[twoside]{article}
\usepackage{fleqn,espcrc2,epsfig}

\newcommand{\AmS}{{\protect\the\textfont2
  A\kern-.1667em\lower.5ex\hbox{M}\kern-.125emS}}

\title{Towards the application of the Maximum Entropy Method to finite temperature
 Upsilon Spectroscopy \thanks{Work supported by the  
European Community TMR Program ERBFMRX-CT97-0122}}

\author{ 
M. Oevers\address{Dept. of Physics and Astronomy,University of Glasgow G12 8QQ, U.K.} with 
C. Davies$^{{\footnotesize{\rm{a}}}}$ and 
J. Shigemitsu\address{The Ohio State University, Columbus, Ohio 43210 }}

\begin{document}
\pagestyle{empty}
\setcounter{topnumber}{1}
\begin{abstract}
%-------------------------------------------------------------------------------
According to the Narnhofer Thirring Theorem \cite{NT} interacting systems 
at finite temperature cannot be described by particles with a sharp dispersion law.
It is therefore mandatory to develop new methods to extract particle masses at 
finite temperature. The Maximum Entropy method offers a path to obtain the 
spectral function of a particle correlation function directly. 
We have implemented the method and tested it with zero
temperature $\Upsilon$ correlation functions obtained from an NRQCD simulation.
Results for different smearing functions are discussed.
\end{abstract}

\maketitle
\section{Introduction}
%-------------------------------------------------------------------------------
With $J/\Psi$-suppression being one of the key probes for the quark gluon plasma
\cite{Satz} a nonperturbative understanding of heavy quarkonia at temperatures
above the transition temperature is important. The extraction of particle masses 
at finite temperature is aggravated on a practical level by the compactified 
Euclidean time direction and on a more fundamental level by the Narnhofer Thirring
theorem. One has to employ methods that do not make any assumptions about the 
spectral structure.
A zero momentum Euclidean current current correlator $G(\tau)$ has the 
following spectral representation \cite{huang}:
\begin{equation}
G(\tau)= \frac{1}{\pi}\int_0^\infty 
\frac{\cosh((\frac{\beta}{2}-\tau)\omega)}
     {\sinh(\frac{\beta}{2}\omega)}\rho(\omega) d\omega
\label{defrho}
\end{equation}
The (temperature dependent) spectral function $\rho(\omega)$ contains all the 
real-time physics information we are interested in. Indeed the Fourier transform 
of the retarded correlator is given in terms of the spectral function as
\[
\tilde{G}_R(\omega)=\frac{1}{\pi}\int_{-\infty}^\infty
\frac{\rho(\omega^\prime)}{\omega+i\epsilon - \omega^\prime} d\omega^\prime
\]
Extraction of $\rho(\omega)$ from Eq.(\ref{defrho}) by inversion is numerically 
an ill-posed problem \cite{qcdtaro} and 
the Maximum Entropy method can be seen as a regularization of this ill-posed 
problem, but has in fact a deeper justification from Bayesian statistics.
%%%%%%%%%%%%%%%%%%%%%%%%%%%%%%%%%%%%%%%%%%%%%%%%%%%%%%%%%%%%%%%%%%%%%%%%%%%%%%%
\section{The Maximum Entropy Method}
%-------------------------------------------------------------------------------
The Maximum Entropy method (MEM) is a well known technique for image 
reconstruction and has been successfully applied in astronomy and condensed 
matter physics. For a review see \cite{JarGub}.
MEM is based on Bayesian methods of inference which in turn
centers around Bayes theorem of conditional probabilities, which in our case reads:
\begin{equation}
P\left[ \rho | G, I \right]  \sim P\left[ G | \rho, I \right]  
                                   P\left[ \rho | I \right]
\label{bayes}
\end{equation}
where $\rho$ is the spectral function, G are the data for the correlator and I is 
any {\it a priori} information that is relevant to the problem. 
$P\left[ G | \rho, I \right]$ is called the likelihood and is proportional to 
$\exp(-L)$ for a large number of measurements with
\[
L=\frac{1}{2}\chi^2=\frac{1}{2}\sum_{i,j}^{N_\tau}
(F(\tau_i)-G(\tau_i)C_{i,j}^{-1}(F(\tau_j)-G(\tau_j))
\]
$C_{i,j}$ being the covariance matrix and 
$F(\tau) = \int K(\tau,\omega)\rho(\omega) d\omega$ is the 'fit function' in terms
of the spectral density $\rho(\omega)$ and the kernel $K(\tau,\omega)$ defined by
Eq(\ref{defrho}).
$P\left[ \rho | I \right]$ is the prior probability for the spectral function.
Using Bayesian lines of thought \cite{Skilling} one can show that the prior 
probability is of entropic form $\exp(-\alpha S)$.
\[
S = \int [\rho(\omega) - m(\omega) - 
\rho(\omega)\log(\frac{ \rho(\omega)}{ m(\omega)})]d\omega 
\]
$m(\omega)$ is a default model with respect to which we measure the entropy of the
spectral function. The freedom to choose a default model can be used to incorporate
further knowledge, but trustable results should not depend on $m(\omega)$. We 
choose $m(\omega) \sim \omega^2$, which is the perturbative high energy form of 
mesonic spectral functions \cite{Japan1,Japan2}. The real and positive parameter 
$\alpha$ controls the relative weight between the entropy and the the likelihood.
In the algorithm used here to determine the spectral function \cite{Bryan} $\alpha$
is eliminated by marginalization. A detailed analysis shows that one can determine
the probability $P\left[ \alpha | G \right]$ from the data. And the final result 
is then given as a weighted average over $\hat{\rho}_\alpha(\omega)$ which is the 
spectral function that maximises 
$Q = \alpha S - L$, i.e. maximises the conditional probability 
$P\left[ \rho | G, I \right]$ in Eq.(\ref{bayes}):
\[
\bar{\rho}(\omega) = \int \hat{\rho}_\alpha(\omega) 
P\left[ \alpha | G \right] d\alpha 
\]
The maximization of $Q$ over the space of $\rho$ makes use of a Singular Value 
Decomposition of the kernel $K(\tau,\omega)$, by expressing $\rho$ in terms of 
the singular vectors of $K$. In this way the algorithm chooses the appropriate 
degrees of freedom of which there are, by construction, always fewer than the number 
of timeslices. MEM therefore makes no assumption about the spectral shape except 
for that which is dictated by the the discretisation of the kernel.
%%%%%%%%%%%%%%%%%%%%%%%%%%%%%%%%%%%%%%%%%%%%%%%%%%%%%%%%%%%%%%%%%%%%%%%%%%%%%%%
\section{Testing MEM with $\Upsilon$ data}
\begin{figure}[h]
\epsfig{file=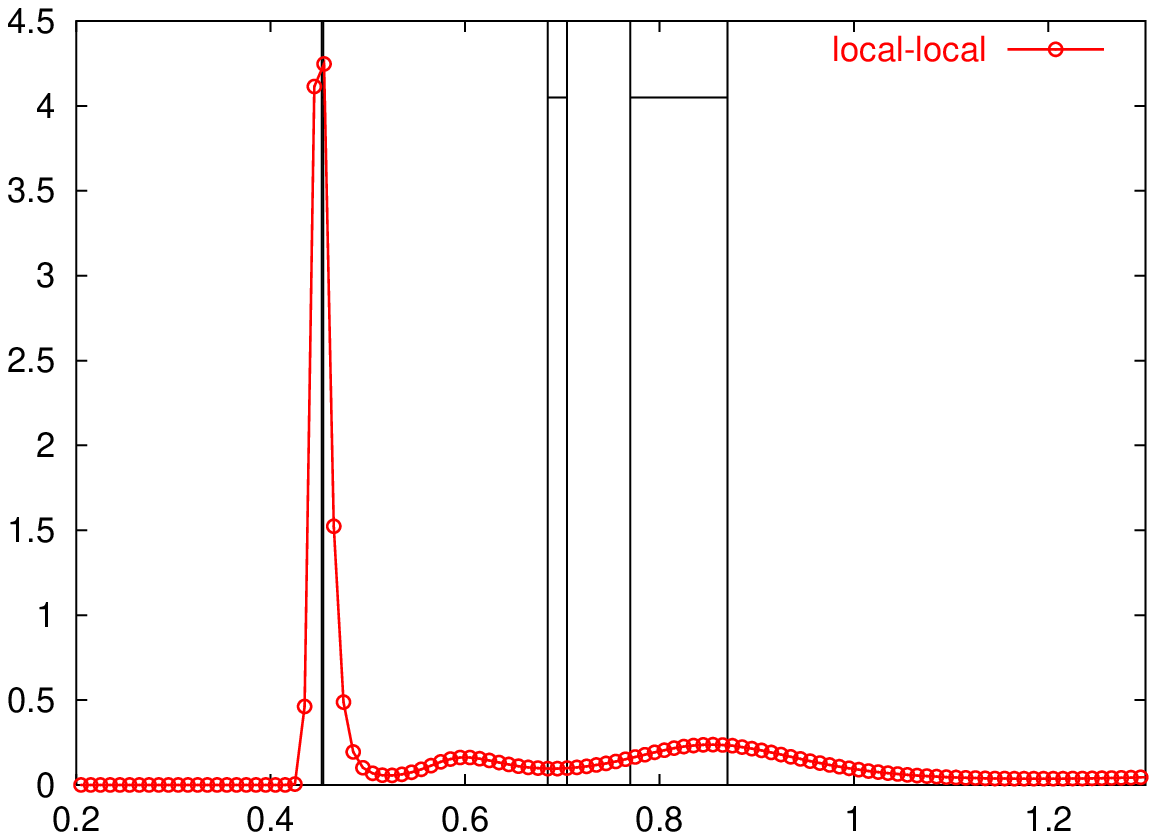, height=0.21\textheight}
\epsfig{file=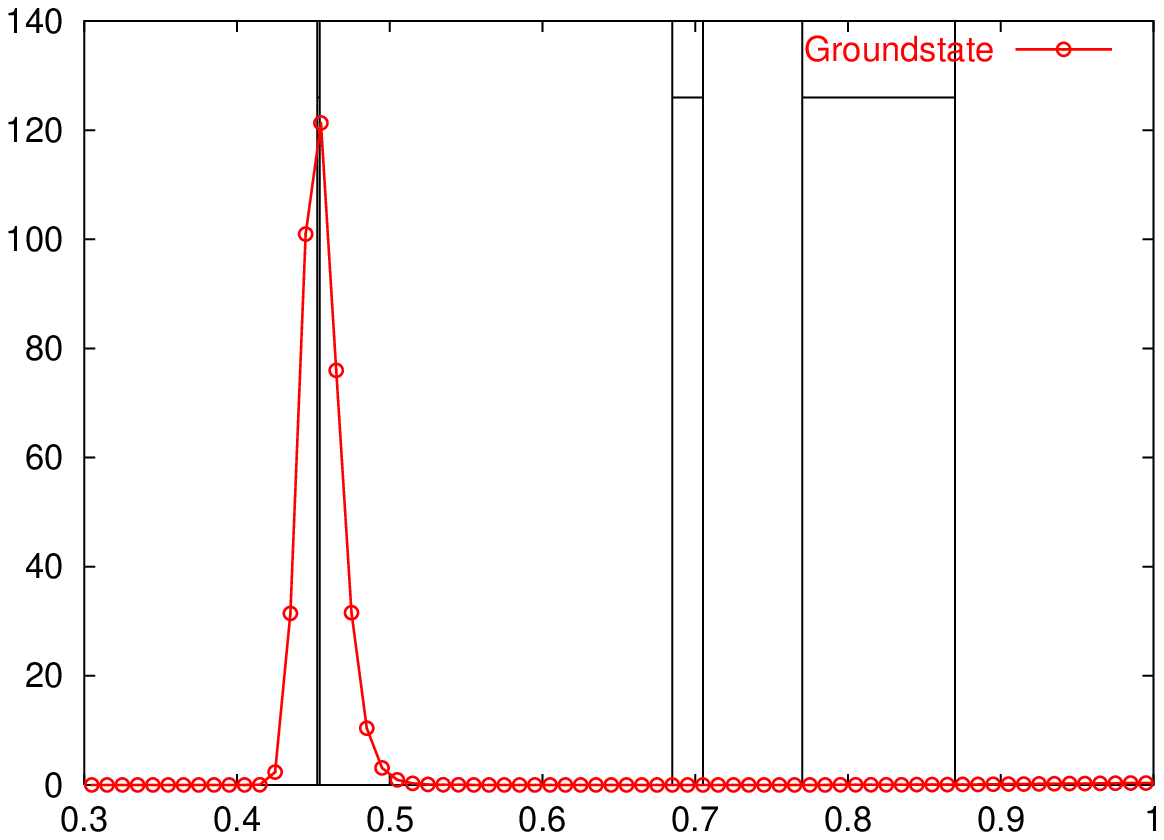, height=0.21\textheight}
\epsfig{file=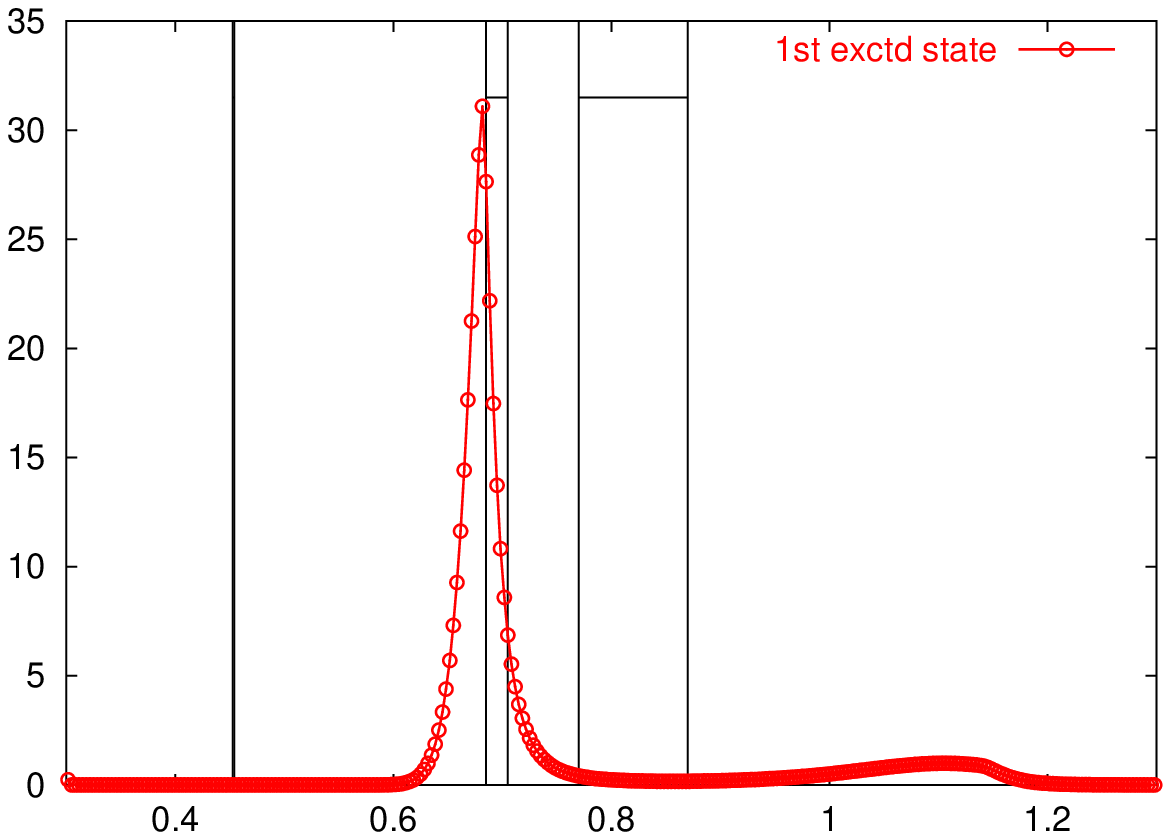, height=0.21\textheight}
\epsfig{file=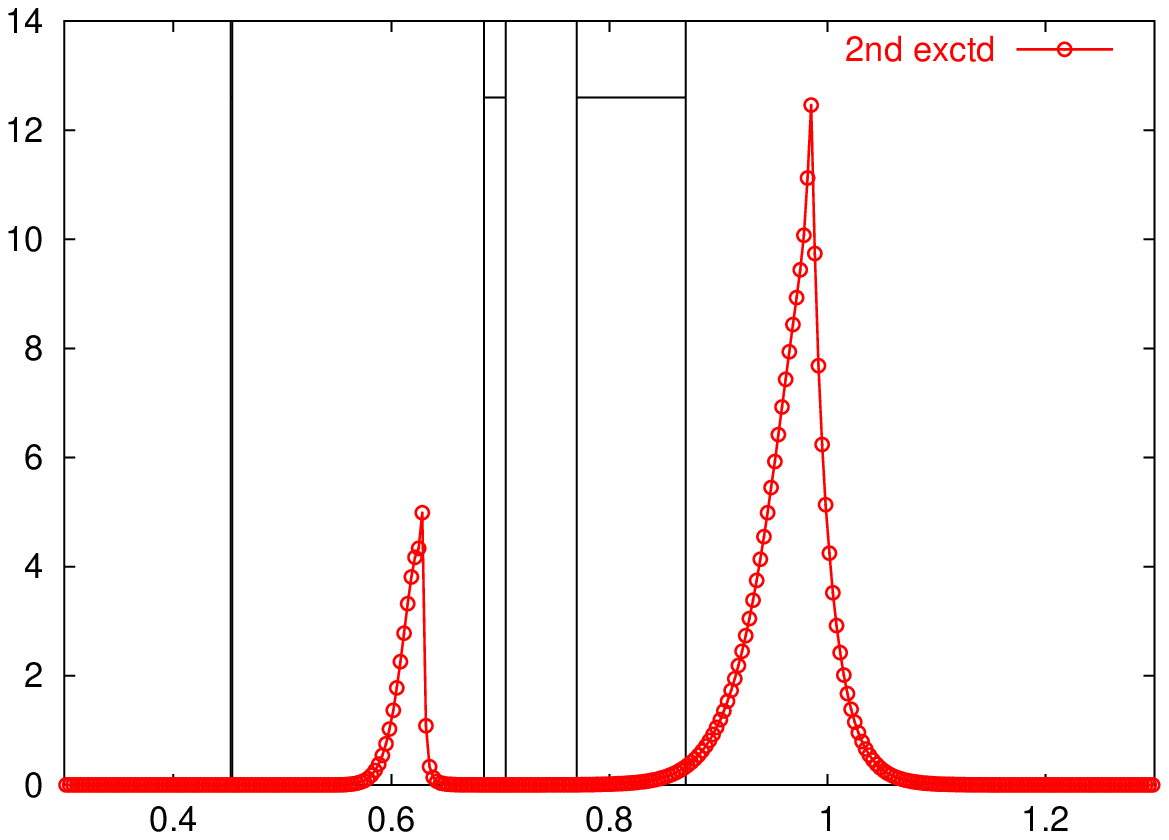, height=0.21\textheight}
%\vspace{-0.1cm}
\caption{Spectral functions $\rho(\omega)/\omega^2$ for (top to bottom) the 
local-local correlator, ground-state smearing, 1st excited state smearing,
2nd excited state smearing. The vertical lines indicate the error-bands of the 
results of \cite{christine}.}
\label{fig:results}
\end{figure}

%-------------------------------------------------------------------------------
Since we are ultimately interested in studying the melting of $\Upsilon$ and 
$J/\Psi$ above the critical temperature, we have tested the method with data
from a precision $\Upsilon$ spectroscopy study at zero temperature using {NRQCD} 
reported in \cite{christine}.
We have analyzed local and smeared correlation functions in the $^3S_1$-channel 
with Richardson potential radial wave functions for the smearings. The smearings
at source and sink have to be identical for Eq.(\ref{defrho}) to hold.  
Fig.\ref{fig:results} shows our results. 
\clearpage
In all graphs we plot 
$f(\omega)=\rho(\omega)/\omega^2$ to divide out the assumed default model 
$m(\omega)=0.1\omega^2$. The integration in Eq.(\ref{defrho}) is performed
up to $\omega_{max}=7.5>2\pi$. The results do not depend on $\omega_{max}$ as 
long as $\omega_{max}>2\pi$, the maximum lattice momentum of the meson. 
We only show the spectral function up to 
$\omega\sim1.2$, since $f(\omega)$ is virtually zero beyond this value except for
$\omega$ close to $\omega_{max}$ when $f(\omega)$ tends to $0.1$, the default model
value. The data do not contain enough information to constrain the spectral 
function at such high momenta.
The results also do not depend on the discretisation 
$d\omega$ as long as $d\omega$ is smaller than the finest structure that is
contained in the data. These results have been produced with $d\omega=0.01$.
We have also checked the independence of our results from the default model.
We have varied the prefactor of $\omega^n$ between 0.002 and 5.0 and the power
$n$ of $\omega^n$ between 0 and 3.
\subsection{Discussion of the results}
%-------------------------------------------------------------------------------
The local-local correlator shows a clear ground-state peak at the location expected
from the analysis in \cite{christine}. There is some indication for spectral 
strength of excited states, but the shape is not very pronounced. Similar behaviour
has been reported in \cite{Japan1,Japan2}. What is new here is the analysis of 
smeared correlators. Although a smeared correlation function probes the same 
quantum numbers as the local correlator, the shape of the spectral function of a 
smeared operator will in general be different from the local spectral function.
However the position of a peak in the spectral function should not be affected
by smearing. The second plot of Fig.\ref{fig:results} shows the result for 
ground-state {smearing} and indeed one finds a clear peak with no indication of any 
excited states. The next plot is the smearing for the first excited state. The 
peak position is consistent with the results from the standard analysis. There is
no contamination from the ground-state and only a slight indication of higher 
excited states. For the last plot a comment is in order. We have not said anything
so far about the statistical significance of our results and if we take the last 
plot at face value it presents us with a problem. The lower of the two peaks lies
lower than the result for the first excited state. Either the 1st excited state 
smearing did in fact project onto a higher state and this is the true first 
excited state or something is problematic with the 2nd excited state smearing.
The second peak which we expect to be the second excited state lies higher than 
the result from the correlated matrix fit of \cite{christine}. The large error 
band from the standard analysis already indicates that this state is difficult to 
analyse. The error analysis described in the next section indicates that both 
peak positions are in fact compatible with the results of \cite{christine} and 
that the data for this smearing are probably not good enough to pin down this 
state. One expects that higher excited states become more and more visible as one 
increases $N_\tau$, the number of points in the time direction. At finite temperature
it will therefore be advisable to use anisotropic lattices in order to have large
$N_\tau > 24 $ at still manageable spatial lattice sizes.
\subsection{Rotated smearings}
%-------------------------------------------------------------------------------
Suppose one has a set of smearings 
{$\left.|0\right>,\left.|1\right>,\left.|2\right>...$}. Measuring all 
cross correlators $\left<0|0\right>,\left<0|1\right>,\left<1|0\right>...$, one
can construct all correlators with smearings that can be expressed as linear 
superpositions of the original smearings: 
$\left.|\Psi\right>=\alpha_0\left.|0\right>+\alpha_1\left.|1\right>+
                    \alpha_2\left.|2\right>+...$.
The correlator $G(\tau)=\left<\Psi(\tau)|\Psi(0)\right>$ will then also have a
spectral representation which can be analysed with our methods. 
Fig.2 is an example of such a procedure.
\vfill
\begin{minipage}{0.48\textwidth}
\epsfig{file=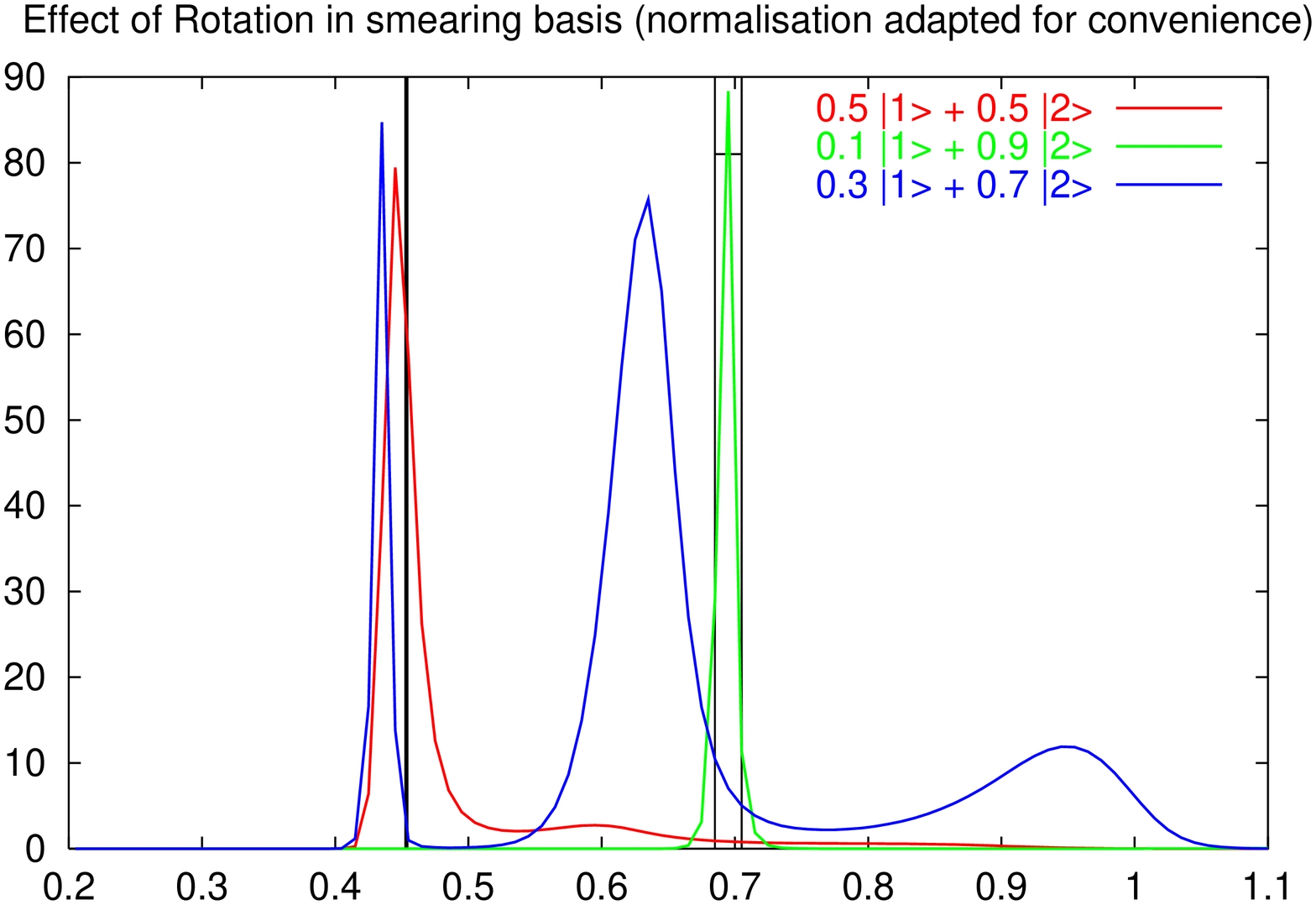, width=0.26\textheight,angle=270}
\end{minipage}\\[4ex]
Figure 2. Spectral functions for 3 different superpositions 
of the ground-state and excited state smearings.
\newpage
This could be used to construct optimised \mbox{smearing} functions that contained little
or no contamination from unwanted states.
\subsection{Error analysis}
%-------------------------------------------------------------------------------
An important issue is of course the error analysis for MEM reconstructed spectral 
functions. One important point to note here is the fact that the spectral 
function is a density. It therefore does not make sense to assign errors to 
individual points, since different points are correlated because e.g. the 
normalization of $\rho$ is fixed. One can adopt an approach suggested in 
\cite{JarGub} in which the covariance of $\rho$ around the maximum of $Q$
is averaged over $\alpha$. This is an approach within the logic of Bayesian 
statistics. We have taken a more reserved approach and asked how robust is the 
MEM prediction under a change of sample. To investigate this we have created
bootstrap samples from the original data and run the whole analysis for every
such sample. For the ground-state smearing this procedure results in a stable 
prediction. The peak position does not change, only the shape of the peak. 
For the 1st excited \mbox{state} smearing the situation is still very good. The result
of this analysis for the 2nd excited \mbox{state} smearing is displayed in Figure 3.
It is clear from this figure that one cannot trust the result for this smearing.
\vfill
\begin{minipage}{0.48\textwidth}
\epsfig{file=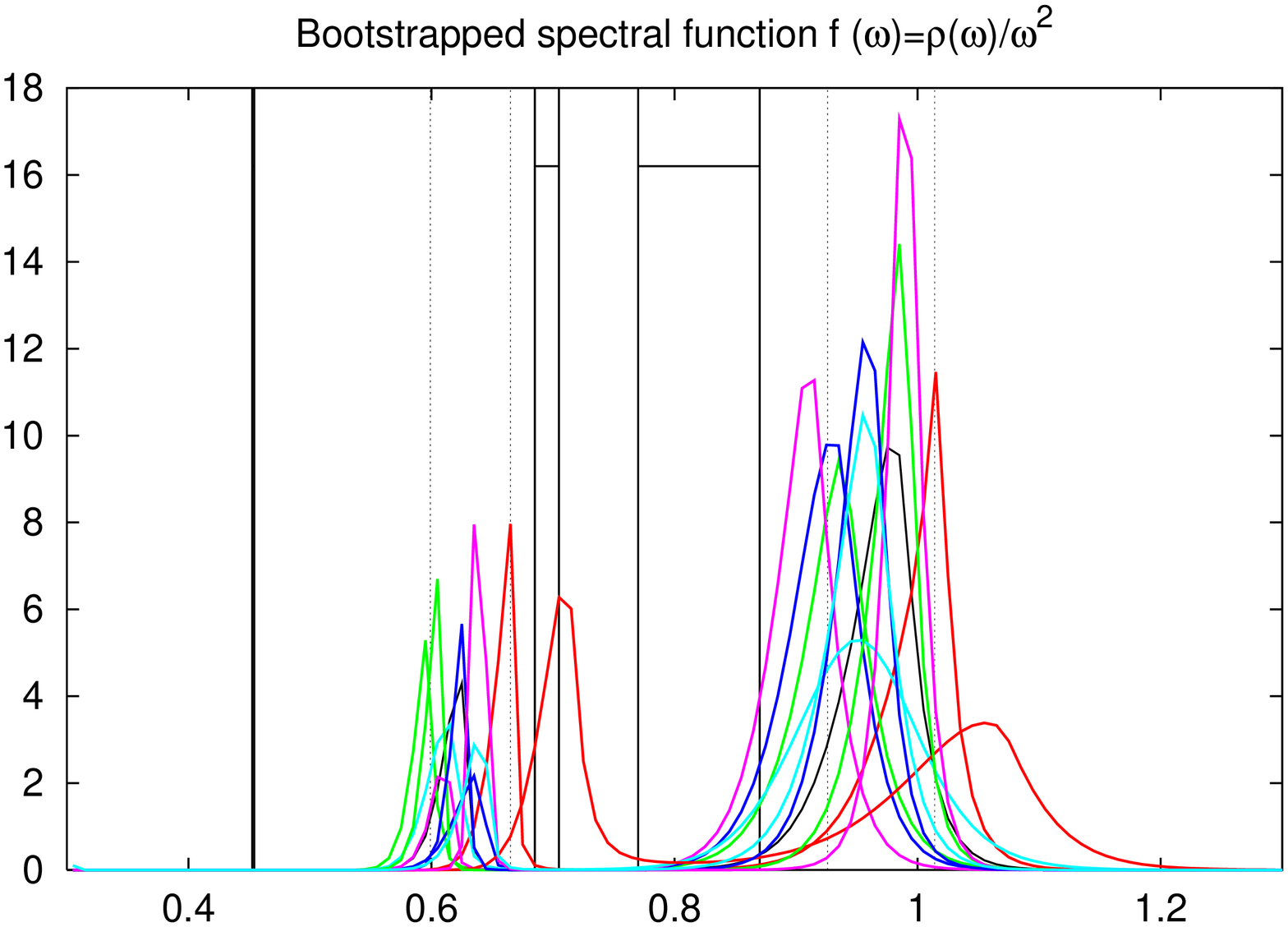, width=0.25\textheight,angle=270}
\end{minipage}\\[0.6cm]
Figure 3. Spectral functions for 10 bootstrap samples for the 2nd excited state
smearing showing the variation in shape and position of the peaks. The dotted lines
are errors of the peak position based on 10 bootstraps.
\newpage
This shows that it is possible to investigate the statistical significance of
MEM prediction by general bootstrap methods.
%%%%%%%%%%%%%%%%%%%%%%%%%%%%%%%%%%%%%%%%%%%%%%%%%%%%%%%%%%%%%%%%%%%%%%%%%%%%%%%
\section{Conclusions}
%------------------------------------------------------------------------------
We have shown that with the Maximum Entropy method one has access to the spectral
function of a current-current correlator which \mbox{contains} all the information about
the excitation spectrum of the theory in the channel represented by this current. 
We have also shown that one has good access to excited states when smeared 
operators are used. In particular rotating correlators in a given basis of smearing
functions, one can produce smearings which contain almost no contamination from 
unwanted states, complementing standard spectroscopy techniques. Since the method 
makes no assumptions about the shape of the spectral function, it can also be used
at finite temperature where little is known about its structure except for the 
absence of single particle $\delta$-function peaks.
We intend to use the method to investigate $\Upsilon$ and $J/\Psi$ melting above 
the critical temperature. 
%%%%%%%%%%%%%%%%%%%%%%%%%%%%%%%%%%%%%%%%%%%%%%%%%%%%%%%%%%%%%%%%%%%%%%%%%%%%%%%
\vfill
%\bibliographystyle{par}
%\bibliography{refs}

\end{document}